\documentclass[reprint,pra,aps,notitlepage,nofootinbib]{revtex4-2}
\usepackage{amsmath}
\usepackage{amssymb}
\usepackage{bm,braket}
\usepackage{float} 
\usepackage{tikz} 
\usepackage{verbatim}
\usetikzlibrary{quantikz} 
\usepackage{dcolumn}
\newcolumntype{d}[1]{D{.}{.}{#1}} 
\newcolumntype{L}[1]{D{.}{#1}{1,4}}

\usepackage{hyperref}
\usepackage{multirow}
\hypersetup{
     colorlinks   = true,
     citecolor    = blue
}
\def\be{\begin{equation}}
\def\ee{\end{equation}}
\def\bea{\begin{eqnarray}}
\def\eea{\end{eqnarray}}
\def\bes{\begin{eqnarray}}
\def\ees{\end{eqnarray}}
\def\bi{\begin{itemize}}
\def\ei{\end{itemize}} % ------- Define Greek Lowercase --------

\usepackage{amsthm}
\theoremstyle{definition}

\newcommand{\tr}{\operatorname{Tr}}
\usepackage{tikz}
\usetikzlibrary{braids}
\begin{document}
\title{Hybrid Quantum-Classical Latent Diffusion Models for Medical Image Generation  } 

\author{K\"ubra Yeter-Aydeniz}\email{kyeteraydeniz@mitre.org}\affiliation{Quantum Information Sciences, Optics, and Imaging Department, The MITRE Corporation, 7515 Colshire Drive, McLean, Virginia 22102-7539, USA}

\author{Nora Bauer \thanks{Corresponding author}}\email{nbauer@mitre.org}\affiliation{Quantum Information Sciences, Optics, and Imaging Department, The MITRE Corporation, 7515 Colshire Drive, McLean, Virginia 22102-7539, USA}

\author{Pranay Jain}\email{pranayjain@mitre.org}\affiliation{Robotics and Autonomous Systems Department, The MITRE Corporation, 7515 Colshire Drive, McLean, Virginia 22102-7539, USA} 

\author{Max Masnick}\email{masnick@mitre.org}\affiliation{Public Health, Environmental and Life Sciences Department, The MITRE Corporation, 7515 Colshire Drive, McLean, Virginia 22102-7539, USA}

% Add others? 

\date{\today}
\begin{abstract}
Generative learning models in medical research are crucial in developing training data for deep learning models and advancing diagnostic tools, but the problem of high-quality, diverse images is an open topic of research. Quantum-enhanced generative models have been proposed and tested in the literature but have been restricted to small problems below the scale of industry relevance. In this paper, we propose quantum-enhanced diffusion and variational autoencoder (VAE) models and test them on the fundus retinal image generation task. In our numerical experiments, the images generated using quantum-enhanced models are of higher quality, with 86\% classified as gradable by external validation compared to 69\% with the classical model, and they match more closely in features to the real image distribution compared to the ones generated using classical diffusion models, even when the classical diffusion models are larger than the quantum model. Additionally, we perform noisy testing to confirm the numerical experiments, finding that quantum-enhanced diffusion model can sometimes produce higher quality images, both in terms of diversity and fidelity, when tested with quantum hardware noise. Our results indicate that quantum diffusion models on current quantum hardware are strong targets for further research on quantum utility in generative modeling for industrially relevant problems. %{\color{red}{I think it would be nice if we add some numerical values}} 

%Generative learning models in medical research are crucial to developing training data for deep learning models and advancing diagnostic tools, but the problem of high quality, diverse images is an open topic of research. Quantum-enhanced generative models have been proposed and tested in the literature, but have been restricted to small problems below the scale of industry relevance. In this paper, we propose and test and quantum-enhanced diffusion model and variational autoencoder (VAE) and test it on the Fundus retinal image generation task. In our numerical experiments, the quantum-generated are of higher quality and match more closely in features to the real image distribution compared to classical diffusion models, even when the classical diffusion models are larger than the quantum model. Additionally, we perform noisy testing on a noise-calibrated hardware simulator to confirm the numerical experiments, indicating that quantum-enhanced diffusion can produce high quality images on current hardware. Our results indicate that quantum utility in generative modeling is currently attainable on current IBM Quantum hardware for industrially relevant problems. 
\end{abstract}
\maketitle 
\twocolumngrid %\onecolumngrid 
\section{Introduction}  
%Literature review. Will organize into paragraphs. Full reference list in lit_review.tex 

Deep learning has the potential to offer unprecedented insights in medical fields, but it requires high-quality training data, which is limited by cost, patient privacy, and limited availability. This necessitates the development of generative models for multidimensional medical data that can produce diverse and realistic training data. One particular example of this need is in the domain of ophthalmology, where eye images, such as retinal fundus images are crucial for diagnostic and research purposes. The problem of generating retinal fundus images has been previously explored using Generative Adversarial Networks (GANs) \cite{Shenkut_2022,AHN2023105289}, but the generated images lack diversity. There are several reasons for mode collapse in GAN-based models, which include the use of Kullback-Leibler (KL) divergence as a loss function, where its asymmetry forces the generator to sacrifice certain modes to maintain training accuracy. In the case of sparse environments, the discriminator accelerates model convergence and leads to vanishing of the gradient of the generator \cite{jiangzhou2024dgrm}. To eliminate the mode collapse problem in GAN-based models, research focused on new loss functions. One example is the Wasserstein GAN (WGAN) \cite{arjovsky2017wassersteingan} that minimizes the Earth-Mover distance. WGAN was then improved by replacement of weight clipping with a gradient penalty \cite{gulrajani2017improved}. 
% (add mode collapse). 
Denoising diffusion probability models (DDPMs) \cite{ho2020} have emerged as an alternative to GANs for image generation. A recent study \cite{M_ller_Franzes_2023} found that, compared to GANs, latent DDPMs were able to generate more diverse images with equal or greater fidelity across three medical image types. Additionally, DDPMs have been applied to conditional image generation \cite{rombach2022} and modality transfer problems (e.g., generating 3D data from 2D data) \cite{chen2024,ouyang2024}. For the case of fundus image generation, various DDPMs have been developed with specialization to the image features, such as the vessel segmentation, to produce highly realistic images \cite{alimanov2023denoisingdiffusionprobabilisticmodel,go2023generationstructurallyrealisticretinal}. 

\begin{figure*}[] 
\includegraphics[width=\textwidth]{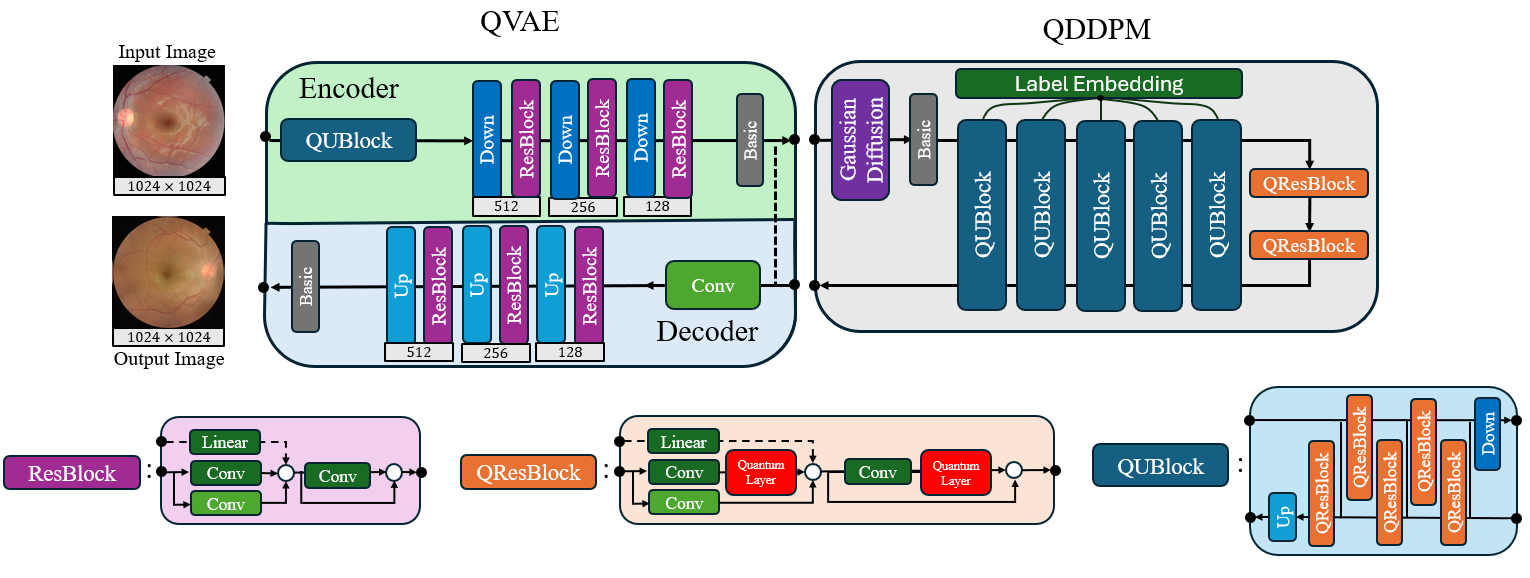}
\caption{Graphical depiction of the QVAE+QDDPM model. The left side shows the QVAE with encoder (green background) and decoder (blue background) structures, and the right side (gray background) depicts the QDDPM. Additionally, structures for the ResBlock, QResBlock, and QUBlock are given in the lower panel. More details on the classical structures can be found in \cite{M_ller_Franzes_2023}. } 
\label{fig:qvae_qddpm}
\end{figure*} 

Quantum generative learning models have been formulated to provide faster training while maintaining stronger data correlations using quantum entanglement and superposition \cite{Amin_2018}. Both fully quantum GANs (QGANs) \cite{Lloyd_2018,Dallaire_Demers_2018} have been introduced, in addition to hybrid quantum-classical models that contain added quantum layers or a fully-quantum generator \cite{Tsang_2023,sekwao2025endtoenddemonstrationquantumgenerative}. These models have been improved with the inclusion of a classical variational autoencoder (VAE), which converts the input data to a representation in latent space \cite{chang2024latentstylebasedquantumgan,chu2025lstmqganscalablenisqgenerative,vieloszynski2024latentqganhybridqganclassical,thomas2024vaeqwganimprovingquantumgans} and applied to various MNIST (Modified National Institute of Standards and Technology)-type datasets to find advantage compared to classical GANs. Recent work has also explored industry-relevant applications, including generating sea route graphs via QGANs \cite{rohe2025investigatingparameterefficiencyhybridqugans} and molecular generation using a hybrid transformer architecture \cite{smaldone2025hybridtransformerarchitecturequantized}. Hybrid quantum-classical GANs have also been studied for medical image generation, particularly for synthetic knee X-ray images generation in \cite{khatun2024quantumgenerativelearninghighresolution}. In \cite{khatun2024quantumgenerativelearninghighresolution} the authors introduced a quantum image generative learning (QIGL) tool, which is a quantum GAN model with a quantum generator that has sub-generators for scalability and a classical discriminator. Principal component analysis (PCA) was utilized to reduce the number of features in the images. QIGL generates knee X-ray images better than classical WGAN; however, use of PCA limits the quality of the generated images, and it is challenging to generalize the model to colored images.  

%\kya{A Hybrid Transformer Architecture with a Quantized Self-Attention Mechanism Applied to Molecular Generation https://arxiv.org/pdf/2502.19214} \cite{smaldone2025hybridtransformerarchitecturequantized} 

Thanks to the benefits in diversity and fidelity of diffusion-based models,  quantum-enhanced DDPM models have been developed for various applications in generating quantum data \cite{Zhang_2024,chen2024quantumgenerativediffusionmodel,kwun2024mixedstatequantumdenoisingdiffusion}, solving ordinary differential equations \cite{wang2025efficientquantumalgorithmsdiffusion},  quantum circuit synthesis \cite{F_rrutter_2024}, and generating MNIST and EuroCon dataset images \cite{De_Falco_2024,cacioppo2023quantumdiffusionmodels}. Quantum diffusion models on MNIST and MedMNIST datasets have been able to have superior generative power as compared to classical diffusion with smaller datasets \cite{chen2025quantumgenerativemodelsimage} or with fewer shots \cite{wang2024quantumdiffusionmodelsfewshot}; however, these studies have not been brought to the large-scale level of industry-relevant applications. Therefore, in this work, we focused on extending the capabilities of quantum diffusion models by integrating them with classical architectures. 

% Not sure if we need further QML discussion or just reference a few of these in Quantum Layer Design 

In this work, we solve an industrially relevant medical image generation problem using a quantum-enhanced DDPM with quantum elements in both the diffusion process and VAE. We confirm using numerical studies that the quantum-enhanced model produces more high-quality images than the classical DDPM, even when the classical DDPM is enhanced by embedding more classical layers. Further, we show using noisy simulations and hardware testing that the quantum-enhanced DDPM is noise resilient and maintains high quality image production when simulated noise is added. Our results indicate that quantum DDPMs are strong candidates for near-term quantum utility for industrial image generation problems. 

Our discussion is organized as follows: in Sec. \ref{sec:methods}, we introduce the quantum-enhanced latent diffusion model and quantum-enhanced variational autoencoder. Additionally, we discuss the choice of quantum layer Ansatz for these constructions and the various metrics we used to determine the optimal Ansatz for the relevant problem and quantum hardware. In Sec. \ref{sec:numerical}, we detail noiseless and noisy numerical simulations of our quantum-enhanced DDPM on Retinal Fundus Multi-disease Image Dataset (RFMID) images and compare with other quantum and classical models. Finally, in \ref{sec:conclusion}, we summarize our outcomes and provide directions for future work. 

\section{Methods}\label{sec:methods} 
Here we introduce the quantum-enhanced latent diffusion model, which in the classical case is generally comprised of a VAE and a DDPM. DDPM models \cite{ho2020denoising, rombach2022high} work by iteratively adding noise to an image, the forwards process, and training the neural network backbone to learn how to remove noise from each noise step, the reverse process. Through this training method DDPMs have showed the promise of generating high quality images from pure noise. The goal of using a VAE in this setup is to provide dimension reduction by providing a compressed latent space in which the diffusion model operates. For completeness, we provide a summary of the classical diffusion model in Appendix \ref{sec:appendixdiff}. The QVAE and QDDPM proposed in this work are based on the classical diffusion model in M\"uller-Franzes, et al. \cite{M_ller_Franzes_2023}, but with added quantum components to enhance the performance by capturing richer correlations and yielding a latent distribution that is more amenable to diffusion thanks to extra expressivity provided by entanglement. 
\subsection{Quantum Variational Autoencoder (QVAE)} 
The quantum variational autoencoder (QVAE) compresses the two-dimensional fundus image space of dimension $1024\times 1024$ into a latent space of size $128\times 128$. Here, we define the quantum residual block (QResBlock), which is a standard residual block (ResBlock) \cite{M_ller_Franzes_2023} enhanced with two quantum layers. The standard ResBlock is obtained when the quantum layers are removed. We also define the quantum UNet Block (QUBlock), which is a standard UNet block \cite{ronneberger_2015} using QResBlocks instead of ResBlocks. This QVAE structure is depicted in the left part of Fig. \ref{fig:qvae_qddpm}, in addition to the ResBlock, QResBlock, and QUBlock structures given in the lower part of the figure. 

The input image is passed through the upper part of the QVAE, the encoder, consisting of the QUBlock, then three layers of Down and ResBlock, which compresses the image from $1024\times 1024$ to $128\times 128$, and then finally a Basic Block which reshapes to the embedding channel dimension. The dashed line indicates that during the training of the QVAE, the data is passed directly from the encoder to the lower part of the QVAE, the decoder. The decoder is comprised of a convolutional block (ConvBlock), three layers of ResBlock and Up, which expands the image from $128\times 128$ to $1024\times 1024$, and a Basic Block which reshapes the output. The QVAE is trained using the Adam optimizer \cite{kingma2017adammethodstochasticoptimization} with a learning rate of 0.001.
\subsection{Quantum DDPM} 
The QDDPM uses a Gaussian diffusion process to gradually corrupt an image with noise in the latent space. A UNet architecture is then trained as a noise estimator such that at each diffusion time step, it predicts the noise that was added. By iteratively removing the estimated noise the UNet ultimately reconstructs or generates synthetic images in the latent space. After passing through the pre-trained QVAE, the data passes through a Gaussian Diffusion process with 1000 steps, through a UNet architecture comprised of a Basic Block to reshape the data, and then through 5 QUBlocks, which contain label embedding in the form of a linear layer. Then, the data passes through 2 QResBlocks and back through the 5 QUBlocks. At this point, the generated image in the latent space is passed through the decoder and the generated image in 2D space is returned. The QUNet architecture is trained using the AdamW optimizer \cite{loshchilov2019decoupledweightdecayregularization} with a learning rate of 0.001. This QDDPM structure is depicted in Fig. \ref{fig:qvae_qddpm}. 
\subsection{Quantum Layer Design} 
We study various variational quantum circuit Ans\"atze designs since the design of the quantum circuit determines the expresssive power, trainability, and quantum hardware efficiency of the studied model. We consider a total of 5 templates for the quantum layer for comparison and each of these templates varies based on the unitary gates utilized and the way the two-qubit entangling operators are implemented. Three are Pennylane \cite{pennylane} templates: Simplified 2 Design (S2D), Basic Entangler (BE), and Strongly Entangling Layers (SE). We also consider two variants of the SE template designed to be more hardware-efficient, edited SE layers variant 1 (ESE) and edited SE layers variant 2 (ESE2). The forms for these Ans\"atze are given in Appendix \ref{sec:appendix_a}. The goal is to test the robustness of these constructions against hardware noise and transpilation on the \texttt{ibm-cleveland} device, which we take to be our target hardware. %{\color{red}{We first say Eagle and then Cleveland}}

\begin{figure*}[ht]
    \centering
    \includegraphics[width=\textwidth]{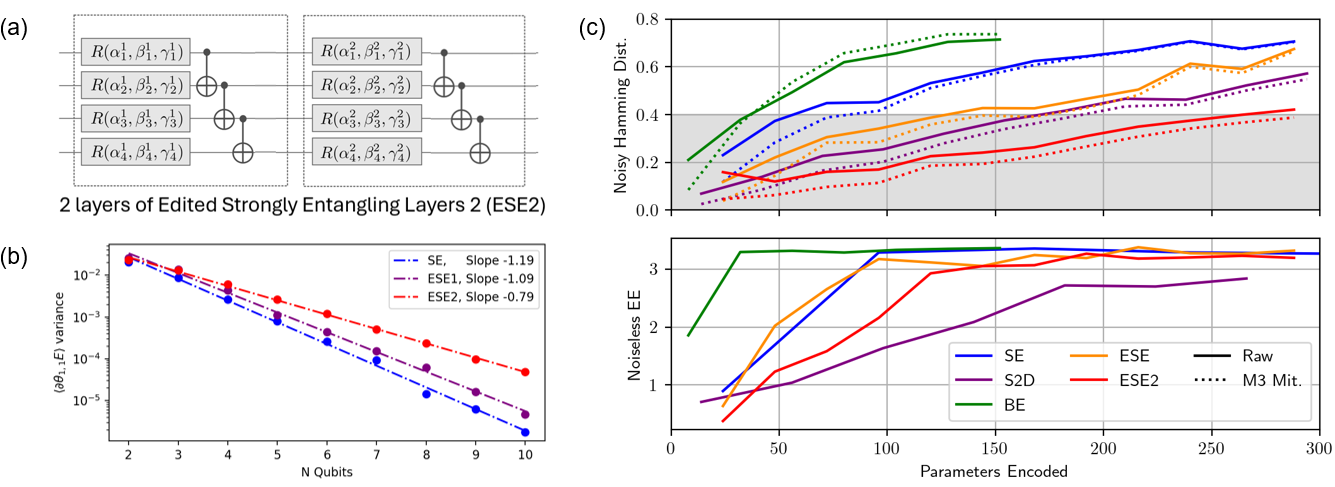}
    \caption{(a) Entangling layer structure for Edited Strongly Entangling Layers 2 (ESE2) Ansatz for a 4-qubit circuit with 2 layers. (b) Average gradient variance $\langle \nabla \theta_{1,1} E\rangle$ for the SE, ESE1, and ESE2 Ans\"atze for 6-layer quantum circuits as a function of the number of qubits. Note that the vertical axis has a Log scale, and the line slopes for each layer are given in the legend as -1.19, -1.09, and -0.79, respectively for each layer studied. (c) Noisy Hamming distance (upper panel) and noiseless entanglement entropy (EE) (lower panel) for the 5 Ans\"atze for 8-qubit circuits considered as a function of the number of parameters encoded. In the upper plot, the raw sampled values are used for the solid lines, and the M3 error mitigated (M3 Mit.) values \cite{mthree} are used for the dotted lines. The gray shading denotes a reasonable threshold for being able to read out the qubit values despite the noise. } 
    \label{fig:ansaetze}
\end{figure*}

We consider specifically the IBM Quantum Cleveland hardware for our simulations. A noise model was constructed in Pennylane \cite{pennylane} that has the same connectivity and one- and two-qubit gate errors for each qubit as an actual device. The main metric considered for quantifying noise and studying the quantum hardware noise robustness is the averaged Hamming distance between the noisy measured bitstrings and the bitstrings sampled from the ideal distribution for randomly sampled circuit parameters. The distance between two sets of samples from the ideal distribution is also calculated as a control to ensure that sufficient samples are taken and to quantify the sampling error. We also consider the entanglement entropy (EE) of the circuits with randomly sampled parameters, given as the von Neumann entropy 
\be S=-\rho \ln \rho~,\ee 
where $\rho$ is the density matrix corresponding to the state produced by the quantum circuit. 
Finally, we utilize the gradient variance (GV) to study the trainability of the circuits with randomly sampled parameters, the scaling of which is used to indicate the tendency of the Ansatz towards barren plateaus. The presence of barren plateaus causes the cost landscape to become flat and it prevents meaningful parameter updates during training. GV is defined as
\begin{equation}
    \text{GV}=\text{Var}_{\bf{\theta}}\left[\nabla_{\theta_i} E({\bf{\theta}})\right]~,
\end{equation}
where $E({\bf{\theta}})$ is the cost function and we used the expectation value of Pauli $Z$ operator as our cost function. 
These metrics as a function of the number of parameters encoded in the Ans\"atze are given in Fig. \ref{fig:ansaetze}. Based on the comprehensive performance of the entangling layer design with respect to the three metrics described, we chose the ESE2 layer design. This design, when faced with the simulated quantum hardware noise, can encode the most parameters with the lowest Hamming distance from the noiseless distribution. The entanglement entropy also plateaus more slowly than the other models; however, while all three-layer designs have an exponentially vanishing gradient variance with qubit number, the gradient variance of the ESE2 layer shrinks at a slower rate of $-0.79$ compared to the SE ($-1.19$) and ESE1 ($-1.09$) layers tested. This indicates that this parameterized quantum circuit will be trainable at a larger size than the other two-layer designs.

Here, we also consider a basic measurement error mitigation protocol, which uses the confusion matrix method to correct for the false positive and negative measurement error rates measured  on the device. This is implemented using the M3 package \cite{mthree}. This has a slight impact by lowering the noisy Hamming distance for the various layer designs. %{\color{red}{We should talk about M3 mitigation and what it means here}}

\section{Numerical Results}\label{sec:numerical} 
In our numerical study, we use the RFMID \cite{rfmid_dataset,data8020029}, which is publicly available. The dataset contains 3200 images divided into training (1920 images), validation (640 images), and testing sets (640 images). Ground truth class labels are provided, including normal/healthy Class 0 and 45 other classes with different types of diseases indicators. We use classes 0, 1, and 2 in our experiments since they are the largest classes. 

We then used our hybrid quantum-classical and classical models to synthetically generate fundus eye images and evaluated the performance of models using commonly used metrics. These metrics are Frechet Inception Distance (FID) score, Centered Maximum Mean Discrepancy (CMMD) score \cite{jayasumana2024rethinkingfidbetterevaluation}, recall, precision, Inception Score (IS), and Automorph \cite{zhou2022automorph} grading results. 

The FID score estimates the distribution gap between the images generated by the model and the distribution of real images. It is defined as 
\begin{equation}
\text{FID}(x, g) = \|\mu_x - \mu_g\|_2^2 + \tr\left(\Sigma_x + \Sigma_g - 2(\Sigma_x \Sigma_g)^{1/2}\right)~,
\end{equation}
where $(\mu_x, \Sigma_x)$ and $(\mu_g, \Sigma_g)$ are the mean and covariance of the feature vectors for the real and generated images, respectively. $\tr(\cdot)$ indicates the matrix trace. To extract the feature vectors of the each image in the dataset, a pre-trained InceptionV3 model that was trained on the ImageNet dataset is utilized. A smaller FID score indicates increased quality and diversity of the generated image dataset. The real images used are a set of evaluation images that were not used in the training or validation. 

Due to limitations of FID score, in \cite{jayasumana2024rethinkingfidbetterevaluation} the authors proposed using CMMD score instead of FID score in evaluating generated images. The CMMD score is a statistical distance metric to measure the similarity between two probability distributions in a feature space. Given that the probability distance of the generated and real images $P$ and $Q$ are over $\mathcal{R}^d$, the CMMD metric with respect to a positive kernel $k$ is defined as 
\begin{equation}
\begin{split}
    \text{CMMD}(P,Q)=&\mathbb{E}_{{\bf{x}},{\bf{x'}}}\left[k({\bf{x}},{\bf{x'}})\right]+\mathbb{E}_{{\bf{y}},{\bf{y'}}}\left[k({\bf{y}},{\bf{y'}})\right]\\&-2 \mathbb{E}_{{\bf{x}},{\bf{y}}}\left[k({\bf{x}},{\bf{y}})\right]~,
    \end{split}
\end{equation}
where ${\bf{x}}$ (${\bf{y}}$) and ${\bf{x}'}$ (${\bf{y}'}$) are independently distributed by $P$ ($Q$). Similar to \cite{jayasumana2024rethinkingfidbetterevaluation} we also used the Gaussian Radial Basis Function (RBF) kernel $k({\bf{x}},{\bf{y}})=\exp\left(-||{\bf{x}}-{\bf{y}}||^2/(2\sigma^2)\right)$ with the bandwidth parameter set
to $\sigma=10$ and CLIP embedding model. To calculate the CMMD metric, we utilized the publicly available code in \cite{cmmd2024} .

The precision metric is the probability that a random generated image falls within the support of the real image distribution, and the recall metric is the probability that a random real image falls within the support of the generated image distribution. 

IS is another commonly used metric, which evaluates the ability of a model to represent the entire ImageNet class distribution. It is defined as
\begin{equation}
    \text{IS}=\exp \left( \mathbb{E}_{x \sim P_g} \left[ D_{KL} \left( p(y|x) \, \| \, p(y) \right) \right] \right)~. 
\end{equation}

The Structural Similarity Index Measure (SSIM) determines the similarity between two image distributions explicitly considering factors such as luminance, contrast, and structure comparison \cite{Wang_2004}, defined as 
\be \mathrm{SSIM} (\mathbf{x}, \mathbf{y})=\frac{(2\mu_x\mu_y+C_1)(2\sigma_{xy}+C_2)}{(\mu_x^2+\mu_y^2+C_1)(\sigma_x^2+\sigma_y^2+C_2)}~,\ee 
where $\mu_i$ is the mean intensity of the image distribution $i$, $\sigma_i$ is the standard deviation, and $\sigma_{ij}$ is the correlation coefficient between image distributions $i$ and $j$. For the coefficients, we use $C_1=(0.01R)^2$ and $C_2=(0.03R)^2$, where $R$ is the range of the image pixel values. 

Automorph is a deep learning pipeline, which initially grades the fundus images according to eye features, and labels the generated images as ``Gradable'' or ``Ungradable''. Then, Automorph computes 72 metrics that describe anatomical features of the fundus images for the set of ``Gradable" images. We generally use 1000 generated images and 1000 real images for the computation of these metrics. The class distribution of the generated images is set to match the proportions of the real image set. The generated images use 100 steps in the sampling process. One notable limitation of Automorph for our purposes is that Automorph was designed in order to assess the gradability for retinal feature measurement on real images, rather than strictly a synthetic image validation tool. We consider Automorph a proxy for a retinal image quality metric. Thus, in our study, we include a variety of metrics to determine the quality of the produced images. 
\subsection{Noiseless Simulation Results} 
In this section, we present the results of both classical and quantum diffusion models, as well as classical and quantum variational autoencoders (VAEs). We compare the performance of 4 models: classical diffusion with classical VAE (CD+CVAE), quantum diffusion with classical VAE (QD+CVAE), classical diffusion with quantum VAE (CD+QVAE), and quantum diffusion with quantum VAE (QD+QVAE). In addition, since the structure of the quantum diffusion model is a hybrid quantum-classical diffusion model with added quantum layers, we also add a classical diffusion model with an additional classical neural network (CDCNN) layer to facilitate a fair comparison in model size. The quantum diffusion layer has $3\times$(\# layers)$\times$(\# qubits) additional parameters, and the CDCNN layer has $4\times$(\# nodes)$^4$ additional parameters compared to the original CD+CVAE model. For the 12-qubit 6-layer model considered here, we have 216 added quantum parameters and 16,384 %82,944 
added CDCNN parameters. % (here, we used 8 qubits {\color{red}{why qubits in classical model?}} for the CNN). 
This overshoot in the number of parameters by $75\times$ is to indicate that the benefit of the quantum layers extends beyond the number of additional parameters and that the quantum entanglement resource can be more effective than classical nonlinear layers even with fewer added model parameters. Finally, we include a self-evaluation between the training and testing images to provide a reasonable baseline for the sample size. 

The summary of the model performance metrics is given in Table\ \ref{tab:diffvae_results}. In these results, the quantum-enhanced models were run on classical simulation of quantum circuits with no sampling or quantum hardware noise included. The optimal FID score is obtained by the CDCNN+CVAE model, and the optimal precision and CMMD are obtained by the CD+CVAE model. QD+CVAE has the optimal recall and IS, and CD+QVAE has the optimal SSIM value. QD+QVAE produces the highest percentage of gradable images, which is a strong indication of model performance as it is an external method of validation. A potential interpretation of these results is that QD produces more variety in the images, and QVAE captures more realistic features of the images, such that the combined QD+QVAE model produces images that most closely resemble the real images. It should be noted that while Automorph grades images as ``gradable'' or ``ungradable'', it was designed as a tool for retinal feature measurement and is not strictly for validation. Thus, a thorough assessment of the ability of Automorph to grade generative model quality is necessary, and the optimal method of external validation is a subject of future research. Note that the self-evaluation for Class 2 has only 10.3\% gradable images, as Class 2 images have the media haze condition label, which results in cloudy images that might not be graded as real.

\begin{table*}[]
    \centering
    \begin{tabular}{ |p{2.5cm}||p{1.4cm}|p{1.5cm}|p{1.7cm}|p{1.5cm}|  p{1.0cm}|p{1.25cm}||p{1.8cm}|p{1.8cm}|p{1.8cm}| }
    \hline 
Model & FID $\downarrow$ & CMMD $\downarrow$ & Precision $\uparrow$ & Recall $\uparrow$ & IS $\uparrow$ & SSIM $\uparrow$ & \% Gradable Class 0 $\uparrow$ & \% Gradable Class 1 $\uparrow$ & \% Gradable Class 2 $\uparrow$ \\ \hline \hline 
CD+CVAE & 32.809 & \textbf{0.091} & \textbf{0.824}	& 0.2050	& 1.6191 & 0.718 & 68.9 & 66.4 & 60.9 \\ \hline % old results 41.52 & \textbf{0.836} & 0.268 & 1.623 & 42.27 & 31.76 & 20.69 \\ \hline 
%QD+CVAE old & 68.165 & 0.302 & 0.673 & \textbf{0.377} & \textbf{1.764} &  & 9.28 & 10.34 & 5.20 \\ \hline 
% more recent QD+CVAE & 50.399 & 0.302 & 0.756 & \textbf{0.259} & \textbf{1.633} & 0.721 & 36.1 & 23.2 & 11.8 \\ \hline 
QD+CVAE & 46.553 & 0.186 & 0.705 & \textbf{0.224} & \textbf{1.725} & 0.720 & 51.4 & 38.4 & 28.0 \\ \hline 
% QD+CVAE new &  &  &  &  &  & 29.8 & 32.8 & 12.5 \\ \hline 
CD+QVAE & 30.732 & 0.132 & 0.733 & 0.203 & 1.524 & \textbf{0.733} & 78.5 & 72.9 & 60 \\ \hline % & 41.13 & 36.87 & 17.68\\ \hline 
QD+QVAE & 29.610 & 0.218 & 0.648 & 0.176& 1.563 & 0.724 & \textbf{85.9} &  \textbf{85.6} & \textbf{74.4} \\ \hline 
CDCNN+CVAE & \textbf{29.513} & 0.112 & 0.786 & 0.196 & 1.605 & 0.728 & 68.8 & 51.2 & 47.3 \\ \hline \hline 
Self & 12.768  & 0.043 & 0.791 & 0.725 & 0.729 & 0.756 & 98.6 & 77.4 & 10.3* \\ \hline 

    \end{tabular}
    \caption{Results for the studied metrics (FID, CMMD, Precision, Recall, IS, SSIM, and Automorph grading for Class 0, Class 1, and Class 2 images) for 5 model types (CD+CVAE, QD+CVAE, CD+QVAE, QD+QVAE, CDCNN+CVAE, respectively), and self-evaluation based on 1000 sampled images. QD+QVAE model outperforms other models in generating gradable images based on Automorph grading.  } 
    \label{tab:diffvae_results}
\end{table*}

%\textcolor{red}{Add SSIM, PSNR, Cosine Similarity. Test for better CDCVAE and QDCVAE points. } 
\begin{figure}[]
    \centering
    \includegraphics[width=\linewidth]{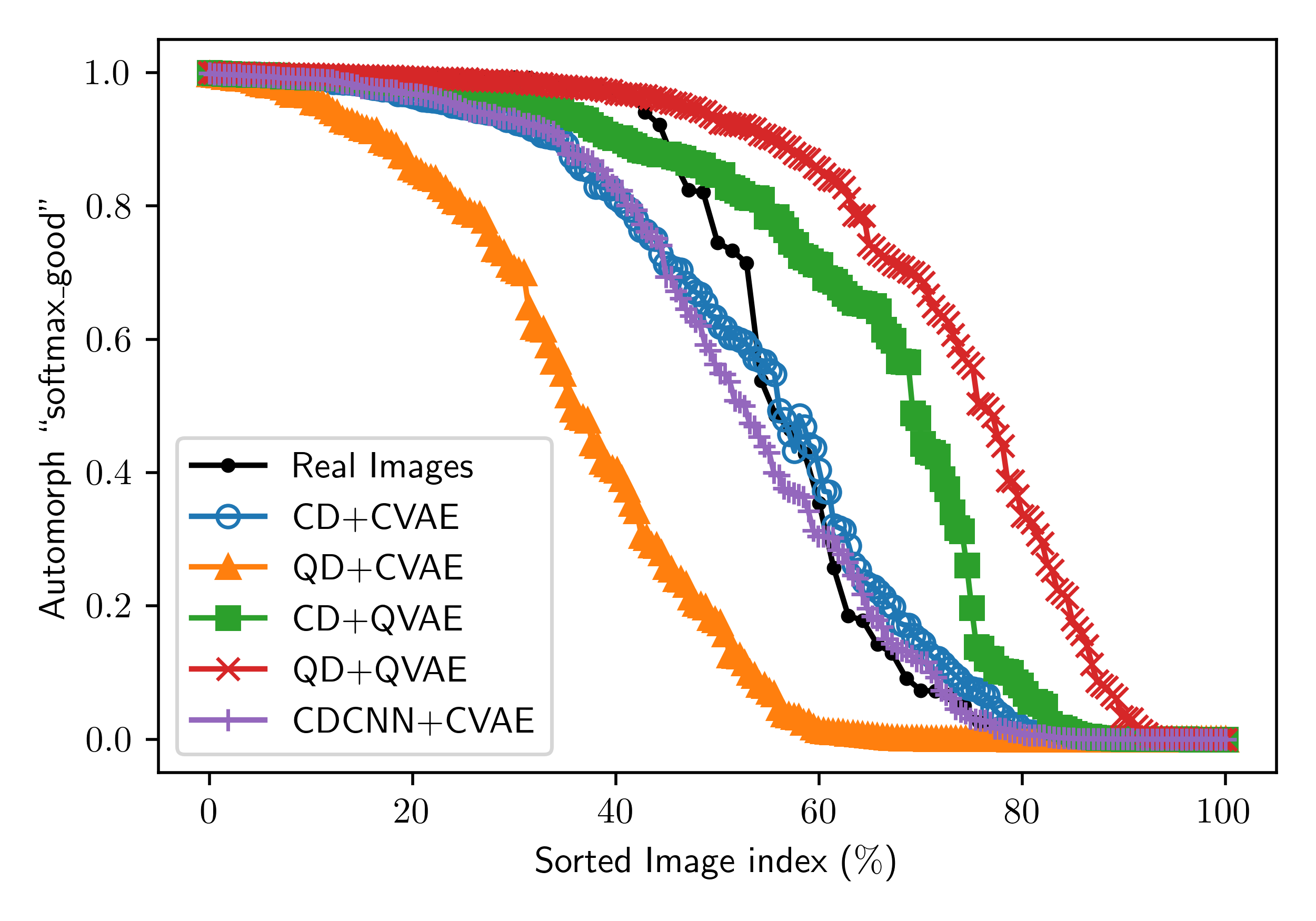} 
    \caption{The Automorph ``softmax good'' metric for 208 Class 0 images generated by the 5 models (CD+CVAE, QD+CVAE, CD+QVAE, QD+QVAE, CDCNN+CVAE) and the real, Class 0 images sorted by value.   }
    \label{fig:automorph_softmax}
\end{figure}

In addition to the gradable or ungradable classification by Automorph, we can also compute the confidence level that an image is classified as ``good'', indicated by the ``softmax good'' metric. The ideal value is 1, while a value of at least 0.5 indicates that the image may be classified as good. We consider 208 images from Class 0 from the 5 categories as well as the testing images and compute the distribution of this confidence value, given in Fig.\ \ref{fig:automorph_softmax}. The real images have over 50\% of the images within the 0.95 confidence interval, which is higher than any of the generated models give. Over 80\% of the QD+QVAE images have $>0.5$ confidence, which is the highest of the models tested. Additionally, the performance of the CD+CVAE and CDCNN+CVAE are very similar, indicating that additional classical layers do not obviously result in a performance improvement, and that the performance is not strictly limited by classical model size. Interestingly, utilizing quantum layers in the diffusion model only (denoted as QD+CVAE) did not improve the performance of the model in terms of most of the metrics, such as FID and SSIM scores or Automorph grading; however, it indicates a slight improvement in the variety of the generated images measured by recall and IS metrics. % QD+CVAE has worse performance than the other three models, but a higher variety, as indicated by the high recall score. 

Further, Automorph also uses a deep learning method to measure 72 features of the vascular structure, disk, and cup of the fundus images. Here, we select the top 20 images as graded by Automorph and run this full feature measurement for the real images and the 5 considered models. We compute the average normalized mean squared error (MSE) between the real distribution and the generated distribution over all 72 features (All), the retinal disc/cup region (Disc/Cup), vascular features of the entire retina (Vascular metrics), and the vascular features restricted to Zones B (C) (Vascular metrics B (C)) of the retina. The disc/cup region includes 6 metrics: disc and cup heights and widths, and the cup-to-disc ratios (CDR) for the height and width. The vascular features include the fractal dimension, vessel density, width, and tortuosity density (three metrics each for veins, arteries, and total). Vascular metrics B and C calculate the vascular feature metrics described, but are restricted to Zones B and C of the retina, respectively. More details on these metrics can be found in Zhou, et al. \ \cite{zhou2022automorph}. A summary of these MSE results is given in Table\ \ref{tab:automse}. 

Overall, the QD+QVAE images match the real image distribution most closely with the lowest MSE. CD+QVAE had the best performance in the disc/cup area, but QD+QVAE had the best performance for the whole image vascular metrics and the metrics restricted to both Zones B and C. QD+CVAE has the best performance in Zone B. These results are a strong indication that the QVAE has superior performance to the CVAE in resolving fine details of the eye images, since the overall best results were with the QVAE models.

\begin{table}[] 
    \centering
    \begin{tabular}{|p{1.75cm}||c|c|p{1.25cm}|p{1.4cm}|p{1.4cm}|} 
    \hline 
	& All	& Disc/Cup	& Vascular Metrics	& Vascular Metrics B	& Vascular Metrics C \\ \hline \hline 
	
CD+CVAE	& 0.1664 & 0.1587 & 0.1345 & 0.2057 & 0.1546 \\ \hline 
QD+CVAE	& 0.1942 & 0.1584 & 0.2049 & \textbf{0.1293} & 0.109 \\ \hline 
%& 0.1882 & 0.1573 & 0.1659 & 0.2327 & 0.1714 \\ \hline 
%CD+QVAE	& 0.2165	& 0.1832	& 0.3251	& \textbf{0.1043}	& \textbf{0.087} \\ \hline 
CD+QVAE	& 0.1675 & \textbf{0.139} & 0.12 & 0.2241 & 0.1574 \\ \hline 
QD+QVAE	& \textbf{0.1361} & 0.1484 & \textbf{0.1169} & 0.1754 & \textbf{0.1077} \\ \hline 
CDCNN +CVAE	& 0.1761 & 0.1415 & 0.1235 & 0.2344 & 0.1703 \\ \hline 
    \end{tabular}
    \caption{Average normalized mean squared error for the various eye metric categories computed by Automorph, comparing the generated image to the real image distribution for models CD+CVAE, QD+CVAE, CD+QVAE, QD+QVAE, CDCNN+CVAE, respectively. Categories include overall metrics (All), optic disc/cup features (Disc/Cup), vascular features (Vascular metrics), and vascular features restricted to Zones B and C. The QD+QVAE model achieves the lowest overall MSE, excelling in vascular metrics, while CD+QVAE performs best in the Disc/Cup category and QD+CVAE in Zone B metrics. }
    \label{tab:automse}
\end{table}

% Additional results for direct comparison between the CVAE and QVAE are given in Appendix \ref{sec:appendix_b}. 
A sample of real fundus images and generated images using CDCNN+CVAE, CD+QVAE, QD+CVAE, QD+QVAE models, respectively are provided in Appendix \ref{sec:appendix_b}. 
\subsection{Noisy Simulation Results} 
Here, we perform noisy simulations using readout and statistical noise. While quantum noise is inherent to currently available quantum devices and must be considered in near-term algorithms, there is also potential for quantum noise to be advantageous in the diffusion model itself. Previous work \cite{Parigi_2024} has proposed quantum noise-based generative diffusion models as a method to obtain more complex probability distributions for sampling. 

First, we switch to a shot-based simulator, where the expectation values are computed from quantum executions with only 1000 shots. Previously, the numerical experiments used statevector simulations, which produced exact expectation values without statistical error. 
Then, we consider the case where expectation values are computed with 1000 shots and a 2.5\%, 5\%, and 10\% readout error probability. The readout error occurs when qubits in the 0 (1) states are incorrectly measured as 1 (0) with the given probability. %Finally, we use a noisy hardware simulator, with error rates calibrated on the IBM Cleveland device. 
The results for the noisy numerical experiments for 8- and 12-qubit models for the studied metrics are given in Table \ref{tab:noisy_results}. 
%Note that for the $12$ qubit model, we were unable to compute a significant number of images on the IBMQ device simulator due to computational constraints by the noisy density matrix simulation, so no values for the studied metrics are given. 
\begin{table*}[]
    \centering
    \begin{tabular}{ |p{3.10cm}||p{1.5cm}|p{1.5cm}|p{1.7cm}|p{1.5cm}|  p{1.1cm}|p{1.4cm}||p{1.6cm}|p{1.9cm}| }
\hline 
 Quantum Model & FID $\downarrow$ & CMMD $\downarrow$ & Precision $\uparrow$ & Recall $\uparrow$ & IS $\uparrow$ & SSIM $\uparrow$ & \% Good $\uparrow$ & \% Gradable $\uparrow$  \\ \hline \hline 
        %12Q, IBM Simulator & & & & & & & & \\ \hline \hline 
        8Q, Noiseless  & 76.877 & 0.182 & 0.340 & 0.230 & 1.741 & 0.696 & 35 & 45  \\ \hline 
        8Q, Readout 2.5 \%  & \textbf{64.003} & \textbf{0.113} & 0.501 & 0.220 & 1.572 & \textbf{0.705} & 43 & 52  \\ \hline 
        8Q, Readout 5 \%  & 64.291 & 0.137 & \textbf{0.520} & \textbf{0.300} & 1.574 & 0.702 & 47 & 57  \\ \hline 
        8Q, Readout 10 \%  & 66.157 & 0.150 & 0.470 & 0.290 & \textbf{1.578} & 0.704 & \textbf{50} & \textbf{59}  \\ \hline 
        %\textcolor{red}{8Q, IBMQ Simulator}  & 74.998 & 0.355 & 0.450 & 0.03 & 1.557 & 0.658 & 14 & 39  \\ \hline \hline 
        12Q, Noiseless  & 64.375 & 0.227 & 0.410 & \textbf{0.420} & 1.581 & 0.748 & \textbf{83} & \textbf{91}  \\ \hline 
        12Q, Readout 2.5\% & 63.719 & 0.221 & 0.430 & 0.410 & 1.576 & \textbf{0.751} & 77 & 88 \\ \hline 
        12Q, Readout 5\% & \textbf{63.695} & \textbf{0.218} & \textbf{0.440} & 0.410 & 1.561 & \textbf{0.751} & 75 & 89 \\ \hline 
        12Q, Readout 10 \% & 69.433 & 0.346 & 0.420 & \textbf{0.420} & \textbf{1.762} & 0.720 & 22 & 28 \\ \hline 
        
    \end{tabular}
    \caption{Results for the studied metrics for 100 Class 0 images generated using the 12-qubit QD+QVAE model and the 8-qubit QD+CVAE model, each with 6 entangling layers, including noiseless and readout noise $\alpha=0.025,~0.05,~0.1$ with 1000 shots. } 
    \label{tab:noisy_results}
\end{table*} 
For the 8-qubit model, the best performing model in terms of the Automorph grading and IS was the 10\% readout error model, while the 5\% readout error model had the best precision and recall, and the 2.5\% error model had the best FID and CMMD scores. The noiseless model did not outperform the readout error models in any metrics. %The IBMQ simulator results had overall worse performance of the models. \textcolor{red}{Rerunning with better transpilation. } 

\subsection{Small Datasets} 
Many industrial applications of generative modeling are limited by small training and validation datasets, which result in poor model performance. Here, we compare the performance of classical and quantum-enhanced diffusion models using quantum and classical VAE (with classical NN) trained on quarter size RFMID data, which was chosen randomly from the original dataset. 

The results for the studied metrics (same metrics studied in full dataset above) for the quarter dataset are given in Table \ref{tab:small_datasets}. For the classical diffusion results, the QVAE outperforms the CVAE by greater margins than with the full dataset, with significant performance gaps between the CVAE and QVAE FID scores, CMMD, and precision. This suggests that the quantum enhancements in the QVAE model are better able to capture features of the images on smaller training sets even with the CVAE model having more parameters, which is further indication that the quantum entanglement and superposition resources are beneficial for industrially relevant generative modeling applications. 
Quantum diffusion with QVAE outperforms all the results in terms of Automorph grading by a factor of at least $2-3\times$, producing a comparable percentage of gradable images as compared to the full RFMID case. %obtaining results for percent gradable images comparable to the full RFMID case. 

\begin{table*}[]
    \centering
    \begin{tabular}{ |p{2.0cm}||p{1.5cm}|p{1.5cm}|p{1.7cm}|p{1.5cm}|  p{1.3cm}|p{1.4cm}||p{1.6cm}|p{1.9cm}| }
\hline 
Model & FID $\downarrow$ & CMMD $\downarrow$ & Precision $\uparrow$ & Recall $\uparrow$ & IS $\uparrow$ & SSIM $\uparrow$ & \% Good $\uparrow$ & \% Gradable $\uparrow$  \\ \hline \hline 
        %CD+CVAE 100 & 73.675 & 0.574 & 0.420 & \textbf{0.390} & 1.617 & 0.718 & 15 & 24 \\ \hline 
        %CD+QVAE 100 & \textbf{59.658} & 0.413 & \textbf{0.670} & 0.340 & 1.502 & \textbf{0.765} & 16 & 31 \\ \hline 
        %QD+CVAE 100 & 94.535 & 0.588 & 0.130 & 0.220 & \textbf{1.912} & 0.670 & 16 & 20 \\ \hline %75.654 & 0.588 & 0.550 & 0.255 & \textbf{1.912} & 0.673 & 16 & 20 \\ \hline 
        %QD+QVAE 100 & 69.549 & \textbf{0.391} & 0.470 & 0.140 & 1.519 & 0.730 & \textbf{53} & \textbf{62} \\ \hline 
        CD+CVAE & 72.732 & 0.547 & 0.380 & 0.268 & 1.617 & 0.721 & 15 & 24 \\ \hline 
        CD+QVAE & 63.151 & 0.413 & \textbf{0.760} & \textbf{0.291} & 1.502 & \textbf{0.766} & 16 & 31 \\ \hline 
        QD+CVAE & 75.654 & 0.588 & 0.550 & 0.255 & \textbf{1.912} & 0.673 & 16 & 20 \\ \hline 
        QD+QVAE & \textbf{57.281} & \textbf{0.391} & 0.510 & 0.123 & 1.519 & 0.731 & \textbf{53} & \textbf{62} \\ \hline 
        
    \end{tabular}
    \caption{Results for the studied metrics for 100 Class 0 images generated using the classical and quantum-enhanced diffusion models with CVAE and 12-qubit QVAE with 6 entangling layers for the quarter RFMID data. } 
    \label{tab:small_datasets}
\end{table*}

\section{Conclusion}\label{sec:conclusion} 
In this work, we developed a quantum-enhanced diffusion model to produce fundus retina images, and compared the performance to its classical counterparts to establish the benefit of quantum layers in generative machine learning. Our model contains both a quantum VAE and a quantum DDPM with hardware-efficient Ansatz layers. Our numerical results indicate that the QVAE+QDDPM model has the best performance, producing the most gradable images according to Automorph image grading and matching most closely to the feature distribution of the real image set, as shown by the studies across a variety of metrics.  Additionally, we optimized the Ans\"atze for the quantum layers for quantum hardware efficiency, and tested the model with the optimal design on a noisy quantum simulator, where adding quantum noise increased both the quality and diversity of the generated image distribution. Finally, when testing on a reduced-size dataset, the QVAE was able to perform significantly better in all metrics compared to the CVAE in CD runs, which indicates another avenue for quantum advantage in generative modeling: using smaller training datasets to produce higher quality images, which is particularly valuable in many industrially relevant use cases due to data scarcity. 

The next priority is testing the sampling performance on quantum hardware. Additionally, it is crucial to study the model training performance on hardware, since this will enable realizing classically intractable quantum model sizes and quantum speedups. This can be aided via quantum transfer learning \cite{Mari_2020}, which would allow quantum modules to be trained to enhance an existing classical model, cutting down on training time and increasing generalizability. 

In order to push toward more advanced use cases, it is important to test the advantage of quantum enhancement for additional image modalities, such as 3D images \cite{khader2023} or time series data \cite{neumeier2024reliabletrajectorypredictionuncertainty}. Additionally, generative models have been proposed that transfer between two imaging modalities, such as positron emission tomography (PET)  and magnetic resonance imaging (MRI) \cite{sun2019dualglowconditionalflowbasedgenerative}. This is an opportunity to expand the use cases of quantum-enhanced diffusion models beyond synthetic image generation. Work in this direction is in progress. 

%MMGen: Unified Multi-modal Image Generation and Understanding in One Go: \cite{wang2025mmgenunifiedmultimodalimage} 

%Quantum Gradient Class Activation Map for Model Interpretability: \cite{lin2024quantumgradientclassactivation}

%Transfer Learning Analysis of Variational Quantum Circuits: \cite{tseng2025transferlearninganalysisvariational} 

While we used a simple diffusion process in our model, continuous normalizing flows (CNFs) are a more general framework that model arbitrary probability paths, including diffusion paths \cite{song2021maximumlikelihoodtrainingscorebased}. Recent work on the flow matching CNF training approach \cite{lipman2023flowmatchinggenerativemodeling} has found faster and smoother training than standard diffusion frameworks, as in our work. Therefore, in order to use the state of the art classical method, it is a natural step to develop a quantum-enhanced flow matching model. 

%Diffusion Normalizing Flow \cite{zhang2021diffusionnormalizingflow} 

%Hybrid Quantum-Classical Normalizing Flow: \cite{zhang2024hybridquantumclassicalnormalizingflow} 

%Further, another avenue for quantum advantage in generative learning is increased model interpretability. The Quantum Gradient Class Activation Map (QGrad-CAM) \cite{lin2024quantumgradientclassactivation,tseng2025transferlearninganalysisvariational} provides explicit formulas for determining the importance of activation maps within neural networks. The results indicate that QGrad-CAM is able to create significant visual ``explanations" that discriminate between classes. It is pertinent to test these features in our quantum models. {\color{red}{Should we remove this?}}

%Medical Diffusion: Denoising Diffusion Probabilistic Models for 3D Medical Image Generation: \cite{khader2023} 

Finally, the trade-off between the expressivity of a quantum circuit and the absence of barren plateaus presents a barrier toward the scalability of quantum variational algorithms \cite{cerezo2024doesprovableabsencebarren}. In response, quantum reservoir computing (QRC) has been proposed as a barren-plateau-free method, since it involves training the classical output of a quantum circuit or evolution instead of quantum gate parameters \cite{Bravo_2022}. QRC has been realized experimentally in analog quantum hardware using up to 108 qubits \cite{kornjaca2024largescalequantumreservoirlearning}; however, maximizing the expressive power of quantum reservoirs is an ongoing topic of research \cite{Goetting_2023,schutte2025expressivelimitsquantumreservoir}. Therefore, it is imperative to determine and test the utility of adding QRC in generative models such as our quantum-enhanced diffusion model. 

%A Comparative Analysis of Adversarial Robustness for Quantum and Classical Machine Learning Models \cite{wendlinger2024comparativeanalysisadversarialrobustness} 

%Robust in practice: Adversarial attacks on quantum machine learning \cite{Liao_2021} 

%Quantum reservoir computing using arrays of Rydberg atoms - Theoretical proposal for Rydberg atom reservoir computing \cite{Bravo_2022} 

%Large-scale quantum reservoir learning with an analog quantum computer - Experimental results from QuEra for reservoir computing with up to 108 qubits \cite{kornjaca2024largescalequantumreservoirlearning}

%Exploring quantumness in quantum reservoir computing - Effect of dephasing and entanglement on reservoir memory \cite{Goetting_2023} 

%Reliable Trajectory Prediction and Uncertainty Quantification with Conditioned Diffusion Models \cite{neumeier2024reliabletrajectorypredictionuncertainty} 

\section{Acknowledgements} 
The authors were supported through the Independent Research and Development Program at The MITRE Corporation. ©2025 The MITRE Corporation. ALL RIGHTS
RESERVED. Approved for public release. Distribution unlimited PR 25-1959. We thank Jacob Lenz and Robert Long for their invaluable contributions in shaping the early foundations of this project. 

\bibliographystyle{nature} % switched to Nature style reduce long author lists to et. al. for a more realistic page count 

\onecolumngrid
\appendix
\section{Diffusion Model}\label{sec:appendixdiff}
Diffusion models are generative models which are defined through a Markov chain over latent variables $x_1, \dots, x_T$ \cite{khader2023denoising}. We assume that ${\bf{x}}_0$ is an eye fundus image to be generated that follows a certain conditional probability distribution $\chi_c$. The Denoising Diffusion Probabilistic Model (DDPM) is used to learn $\chi_c$. The idea is to transform $\chi_c$ into a standard normal distribution $\mathcal{N}(0,{\bf{I}})$ and then train a neural network to model the reverse process, hence establishing a mapping from $\mathcal{N}(0,{\bf{I}})$ back to $\chi_c$. DDPM consists of two processes, i.e., a forward (diffusion) process, and a reverse (inverse diffusion) process. 

The forward process is a Markov chain that creates transition kernels $q(x_t, x_{t-1})$ to incrementally transform data distributions into tractable prior distributions by adding Gaussian noise as follows.
\begin{equation}
    q({\bf{x}}_t | {\bf{x}}_{t-1}) = \mathcal{N}({\bf{x}}_t; \sqrt{1 - \beta_t} {\bf{x}}_{t-1}, \beta_t {\bf{I}})~,
\end{equation}
where $\beta_t$ controls the noise schedule through fixed variance of the Gaussian noise added at each timestep. Through reparameterization and setting $\alpha_t=1-\beta_t$ \cite{bai2023conditional, liu2025wifi} and $\bar{\alpha}=\prod_{t=1}^T\alpha_i~,$ $x_t$ can be sampled at any arbitrary time step $t$ as
\begin{equation}
    q(x_t|x_0)= \mathcal{N}(x_t, \sqrt{\bar{\alpha}_t}x_0,(1-\bar{\alpha}_t){\bf{I}})~.
\end{equation}
For large numbers of timesteps $T$, we find that $\lim_{T \to \infty}{\bf{x}}_t\sim \mathcal{N}(0,{\bf{I}})~.$ 

Similar to transforming $\chi_c$ into $\mathcal{N}(0,{\bf{I}})$ by adding Gaussian noise, we reverse the process and can map $\mathcal{N}(0,{\bf{I}})$ to $\chi_c$ by removing noise. Traversing the Markov chain in reverse allows for the generation of new images from the learned distribution. The model is trained to minimize the difference between the true noise $\epsilon$ and predicted noise $\epsilon_\theta(x_t,t)$ using a simplified loss $ \mathcal{L} = \mathbb{E}_{t, x_0, \epsilon} \left[ || \epsilon - \epsilon_\theta(x_t, t) ||^2 \right]$.

% The Markov chain is defined as $q(x_t | x_{t-1}) = \mathcal{N}(x_t; \sqrt{1 - \beta_t} x_{t-1}, \beta_t I)$, where $\beta_t$ controls the noise schedule through variance of the Gaussian noise added at each timestep. The initial step of the diffusion model is the noising process in which starting from the original image $x_0$, the image is continuously perturbed by adding Gaussian noise with increased variance for $T$ timesteps. In the denoising process, a neural network is conditioned on the noisy version of the image at timestep $t$ and the timestep is trained to learn the noise distribution employed in the noising process of the image. This process is followed by inferring the data distribution $p(x_{t-1}|x_t)$ at timestep $t-1$. When the number of timesteps $T$ is large enough, the distribution $q(x_T)$ can be approximated by a prior distribution. This allows sampling from this prior distribution, typically normal distribution $\mathcal{N}(0,I)$. Traversing the Markov chain in reverse allows for the generation of new images from the learned distribution. The model is trained to minimize the difference between the true noise $\epsilon$ and predicted noise $\epsilon_\theta(x_t,t)$ using a simplified loss $ \mathcal{L} = \mathbb{E}_{t, x_0, \epsilon} \left[ || \epsilon - \epsilon_\theta(x_t, t) ||^2 \right]$.

To enhance the representation quality and computational efficiency, a variational autoencoder (VAE) \cite{kingma2022autoencodingvariationalbayes} was employed in our diffusion model as a feature extraction mechanism. The VAE architecture encodes high-dimensional image data into a compact latent space which enables the diffusion process to operate in this low-dimensional latent space instead of the original, large image data space. VAEs are also probabilistic generative models which consist of an encoder and a decoder in their architectures. The encoder maps the input data, $x$, into a latent representation $z$ by progressively downsampling the input through convolutional layers, residual blocks, and attention mechanisms. This latent representation produced by VAE is normalized to follow a Gaussian distribution to make it suitable for diffusion modeling. The decoder in VAE reconstructs the input data from the latent space using upsampling layers, residual blocks, and attention mechanisms, similar to the encoder. The quality of image reconstruction is ensured by using the reconstruction loss which consists of the sum of pixel-wise reconstruction loss, structural similarity (SSIM) \cite{wang2003multiscale}, and learned perceptual image patch similarity (LPIPS) \cite{zhang2018unreasonableeffectivenessdeepfeatures}.

\section{Full Ans\"atze }\label{sec:appendix_a}  
\begin{figure*}[]
    \centering
    \includegraphics[width=\textwidth, trim=0 225 0 0, clip]{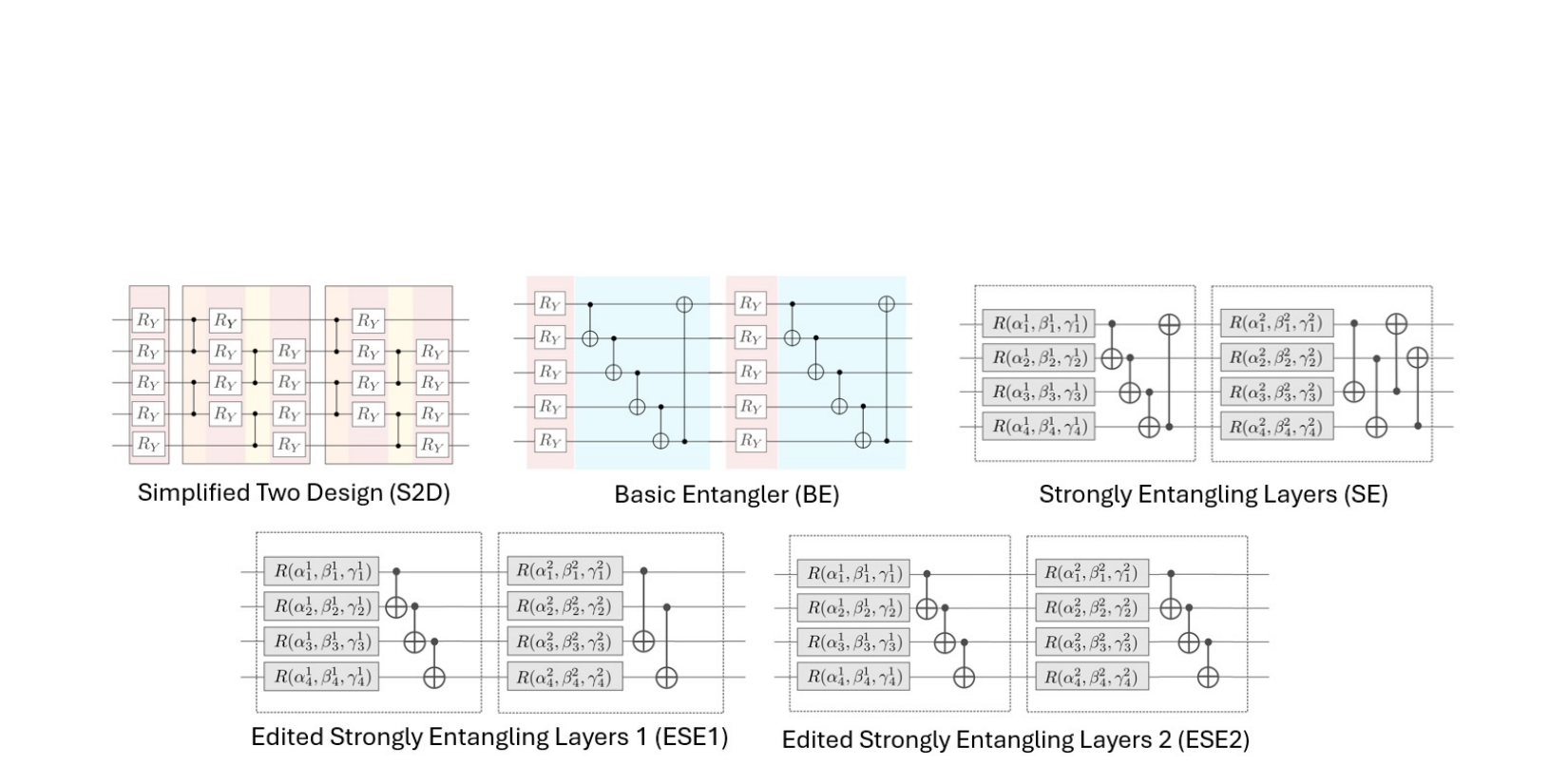}
    \caption{2 example layers of all 5 Ans\"atze (Simplified Two Design (S2D), Basic Entangler (BE), Strongly Entangling Layer (SE), Edited Strongly Entangling Layers 1 (ESE1), and Edited Strongly Entangling Layers 2 (ESE2), respectively) tested in Section \ref{sec:methods} of the main text. The upper row of images have been taken from the Pennylane software \cite{pennylane} website. }
    \label{fig:5ansaetze}
\end{figure*}

Here in Fig. \ref{fig:5ansaetze}, we include the forms of the Ans\"atze tested in Section \ref{sec:methods} of the main text. The first three Ans\"atze are Pennylane \cite{pennylane} templates: Simplified Two Design (S2D), Basic Entangler (BE), and Strongly Entangling Layers (SE). S2D consists of an initial layer of $R_Y$ rotations, then each layer is made up of Controlled-Z ($CZ$) layers between pairs starting with even qubits, an $R_Y$ layer on the affected qubits, and then the pattern is repeated with pairs starting with odd qubits. This Ansatz is of interest in studying barren plateaus via gradient variance scaling as it does not encounter barren plateaus with a local cost function \cite{Cerezo_2021}. BE layers consist of a layer of $R_Y$ gates followed by a ring of Controlled-NOT (CNOT) gates between each qubit of neighboring index, assuming periodic boundaries. SE layers consist of a $U_3$ rotation followed by a ring of CNOT gates between each qubit and its $i$th neighbor (assuming periodic boundaries), where $i$ starts at 1 and increases with each layer until it resets at one less than the total number of qubits. 

In order to improve upon the hardware efficiency of the SE Ansatz, namely, considering the limited qubit connectivity of NISQ (noisy intermediate scale quantum) devices, we introduce ESE1 (Edited Strongly Entangling Layers 1) and ESE2. ESE1 is identical to SE except that it does not use periodic boundaries, and ESE2 is identical to SE except that it does not assume periodic boundaries and keeps $i=1$ at each layer. Quantum circuits with 2 layers of these five Ans\"atze are given in Fig. \ref{fig:5ansaetze}. 
\section{Sampled Images }\label{sec:appendix_b}  
\begin{figure*}[]
    \centering
    \includegraphics[width=\textwidth]{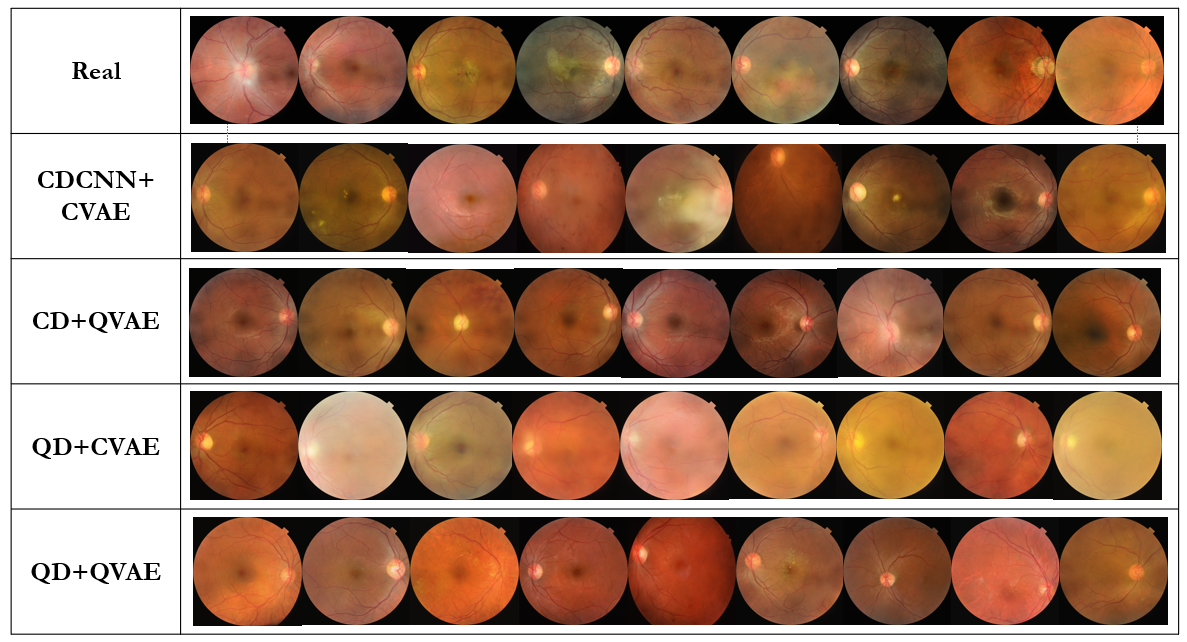}
    \caption{First 9 sampled Class 0 (healthy) images from the real and 4 model types (CDCNN+CVAE, CD+QVAE, QD+CVAE, QD+QVAE, respectively). The QD+QVAE model images have both image variety and distinct features most similar to the real image distribution \cite{rfmid_dataset,data8020029}. } 
    \label{fig:sampled}
\end{figure*}
In this section, we present the first 9 sampled Class 0 images from the real and 4 model types (CDCNN+CVAE, CD+QVAE, QD+CVAE, QD+QVAE, respectively) as seen in Fig. \ref{fig:sampled}. The real images are the target distribution images (taken from the testing set). As indicated by the metrics described in the main text, the CDCNN+CVAE images have high variety but some distortions, which result in the external validation rejecting them as real images. The CD+QVAE images have distinct vessel segmentation and features, but the images have low variety. The QD+CVAE images have high variety but blurry features. The QD+QVAE images have both distinct vessel segmentation and features and higher variety than the CD+QVAE images.

\end{document}